{}
{}

\documentclass[12pt,a4paper]{article}

\newif\ifpublic\publictrue

\newif\ifworking\workingtrue

\usepackage{mathtools,mathrsfs,amsbsy,amssymb,latexsym,amsfonts,amscd,
amsmath}
\usepackage{bm}
\usepackage{xspace}
\usepackage{multirow}
\usepackage{graphicx}
\usepackage[usenames,dvipsnames]{color}
\usepackage[normalem]{ulem}
\usepackage[nosort]{cite}
\definecolor{linkcolor}{rgb}{0,0,0.6}
\usepackage[ pdftex,colorlinks=true,
pdfstartview=FitV,
linkcolor= linkcolor,
citecolor= linkcolor,
urlcolor= linkcolor,
hyperindex=true,
hyperfigures=false]
{hyperref}

\def\showkeysrefformat#1{{\normalfont\tiny\ttfamily#1}}
\makeatletter
\def\SK@@ref#1>#2\SK@{%
{\@inlabelfalse\leavevmode\vbox to\z@{%
\vss\SK@refcolor\rlap{\vrule\raise .75em%
\hbox{\showkeysrefformat{#2}}}}}}
\makeatother

\usepackage[a4paper,text={160mm,247mm},centering]{geometry}
\global\arraycolsep=2pt

\allowdisplaybreaks[3]

\numberwithin{equation}{section}

\expandafter\def\expandafter\bfseries\expandafter{\bfseries\ifmmode\else\boldmath\fi}
\expandafter\def\expandafter\mdseries\expandafter{\mdseries\ifmmode\else\unboldmath\fi}
\expandafter\def\expandafter\normalfont\expandafter{\normalfont\ifmmode\else\unboldmath\fi}

\def\g{\mathfrak g}

\def\h{\mathfrak h}

\def\beq{\begin{equation}}
\def\eeq{\end{equation}}

\def\beqz{\begin{equation*}}
\def\eeqz{\end{equation*}}

\def\bea{\begin{eqnarray}}
\def\eea{\end{eqnarray}}

\def\ha{\mbox{\small $\frac{1}{2}$}}

\def\id{\protect{{1 \kern-.28em {\rm l}}}}

\newcommand{\ti}[1]{_{\bm{{#1}}}}
\def\1{{\ti{1}}}
\def\2{{\ti{2}}}
\def\3{{\ti{3}}}

\newcommand{\dd}{\text{d}}

\def\K{{\mathcal K}}
\def\CC{\mathbb{C}}
\def\PP{\mathbb{P}}
\def\ZZ{\mathbb{Z}}

\def\L{\mathcal{L}}

\newcommand{\LO}{L}

\newcommand{\uucheche}{u}

\newcommand{\ad}{\text{ad}}

\definecolor{lightred}{RGB}{255,127,127}
\definecolor{lightgreen}{RGB}{127,255,127}
\definecolor{lightblue}{RGB}{127,127,255}
\definecolor{linkcolor}{rgb}{0,0,0.6}

\definecolor{brightBlue}{rgb}{0,0,1}

\definecolor{Violet}{rgb}{0.47,0,1}

\usepackage{fancyhdr}

\makeatletter
\let\@keywords\@empty
\let\@subject\@empty
\providecommand{\keywords}[1]{\gdef\@keywords{#1}}
\providecommand{\subject}[1]{\gdef\@subject{#1}}
\def\thetitle{\@title}
\def\theauthor{\@author}
\def\thesubject{\@subject}
\def\thedate{\@date}
\def\thekeywords{\@keywords}
\makeatother
\AtBeginDocument{
\hypersetup{pdftitle={\thetitle}}
\hypersetup{pdfauthor={\theauthor}}
\hypersetup{pdfsubject={\thesubject}}
\hypersetup{pdfkeywords={\thekeywords}}}

\title{Ultralocal Lax connection for para-complex \texorpdfstring{$\mathbb{Z}_T$-cosets}{ZT-cosets}}

\author{F. Delduc, S. Lacroix, M. Magro, B. Vicedo}

\begin{document}

\hfill [ZMP-HH/19-14]
 
\begin{center}

\vspace*{2cm}

\begingroup\Large\bfseries\thetitle\par\endgroup

\vspace{1.5cm}

\begingroup
F. Delduc$^{a,}$\footnote{E-mail:~francois.delduc@ens-lyon.fr},
T. Kameyama$^{b,}$\footnote{E-mail:~takashi.kameyam@gmail.com},
S. Lacroix$^{c,}$\footnote{E-mail:~sylvain.lacroix@desy.de},
M. Magro$^{a,}$\footnote{E-mail:~marc.magro@ens-lyon.fr},
B. Vicedo$^{d,}$\footnote{E-mail:~benoit.vicedo@gmail.com}
\endgroup

\vspace{1cm}

\begingroup
$^a$\it Univ Lyon, Ens de Lyon, Univ Claude Bernard, CNRS, Laboratoire de Physique,
\\
F-69342 Lyon, France,\\

\medskip

$^b$\it 2-10-3, Shimonikura, Wako 351-0111, Japan
\\

\medskip

$^c$\it II. Institut f\"ur Theoretische Physik, Universit\"at Hamburg,
\\
Luruper Chaussee 149, 22761 Hamburg, Germany
\\
Zentrum f\"ur Mathematische Physik, Universit\"at Hamburg, \\
Bundesstrasse 55, 20146 Hamburg, Germany
\\

\medskip

$^d$\it Department of Mathematics, University of York, York YO10 5DD, U.K.\\
\endgroup

\end{center}

\vspace{2cm}

\begin{abstract}
We consider $\sigma$-models on para-complex $\mathbb{Z}_T$-cosets, which are 
 analogues of those on complex homogeneous  
 target spaces  considered recently by D. Bykov. For these models, we show the existence of  
 a gauge-invariant Lax connection 
whose Poisson brackets are ultralocal. Furthermore, its  light-cone components 
commute with one another in the sense of Poisson brackets. 
This extends a result of O. Brodbeck and M. Zagermann 
obtained twenty years ago for hermitian 
symmetric spaces.
\end{abstract}

\newpage

\setcounter{tocdepth}{2}
\tableofcontents
\section{Introduction}

In a classical integrable $(1+1)$-dimensional field theory, the integrals of motion in involution can be extracted from the monodromy of its Lax connection along a constant-time curve. For this reason, the spatial component of the Lax connection, known as the Lax matrix, plays a central role in establishing the property of integrability. In particular, the involution of the integrals of motion is deduced from the specific form of the Poisson brackets of the Lax matrix. In integrable $\sigma$-models the latter are non-ultralocal \cite{Maillet:1985fn,Maillet:1985ek}, in the sense that they
contain a term proportional to the derivative of the Dirac $\delta$-distribution. Yet the presence of such a term
has posed a serious obstacle, for over 30 years, in the problem of quantising such theories from first principles.

Indeed, the most effective and powerful known way  
to quantise a classical 
integrable field theory is to use the Quantum Inverse Scattering Method (QISM) 
\cite{faddtakh1979-1,Kulish:1979if,Faddeev:1979gh}. Unfortunately, the central assumption behind this method is that the Poisson bracket of the Lax matrix of the classical integrable field theory one starts with is ultralocal, \emph{i.e.} does not depend on derivatives of the Dirac $\delta$-distribution.

More precisely, a standard way of applying the QISM is to start by putting the theory on the lattice, which first requires constructing a discretisation of the classical Lax matrix. There are two important properties which such a discretised Lax matrix should have. Firstly, just as in the continuum, one would like its Poisson brackets to have a form which ensures the existence of sufficiently many integrals of motion in involution. A very general family of Poisson algebras with this property is given by the Freidel-Maillet quadratic 
algebras \cite{Freidel:1991jx,Freidel:1991jv}. Secondly, we should also recover the Lax matrix of the field theory from it in the continuum limit. In an ultralocal theory, these two requirements are fulfilled by defining the discretised Lax matrix as the path-ordered exponential of the continuum Lax matrix between two sites. In the non-ultralocal setting, however, the Poisson bracket of the path-ordered exponential of the Lax matrix, on adjacent or overlapping intervals, is not well defined \cite{Maillet:1985fn,Maillet:1985ek,SemenovTianShansky:1995ha} due to the presence of $\delta'$-terms in the Poisson bracket of the Lax matrix.

Faced with the problem of non-ultralocality in any given integrable field theory, it is 
natural   to  seek an alternative Lax matrix for this theory which would 
not suffer from the presence of $\delta'$-terms in its Poisson brackets.   Such an alternative 
has not been found for a generic 
integrable $\sigma$-model. Let us recall that in some cases 
a different  strategy may be applied. It consists in discretising and quantising {\em{\`a la}} 
Faddeev-Reshetikhin. This was first developed for the Principal Chiral Model
 \cite{Faddeev:1985qu} (see  \cite{Appadu:2017bnv,Appadu:2017fff,Appadu:2018jyb} for 
other recent applications  of this approach).   This way of treating non-ultralocality  relies however on 
an ultralocal Lax matrix which is  
associated with a modified canonical structure. 
 
\medskip

Among classical integrable non-linear $\sigma$-models, there are the ones 
on $\mathbb{Z}_T$-cosets  \cite{Pohlmeyer:1975nb,Eichenherr:1979ci,Young:2005jv}.
The Poisson 
brackets of their Lax matrix are non-ultralocal \cite{Sevostyanov:1995hd,Magro:2008dv,
Vicedo:2009sn,Ke:2011zzb}. In this article, we show that classical para-complex 
$\mathbb{Z}_T$-cosets  
also admit  an ultralocal Lax connection.  Moreover, the light-cone components of this Lax connection Poisson 
commute  with one another.  These results generalise the ones obtained in 
\cite{Bytsko:1994ae,Bazhanov:2017nzh}  for the O(3) non-linear $\sigma$-model and 
 in \cite{Brodbeck:1999ib} for hermitian symmetric space $\sigma$-models.
 The complex structure of the latter target spaces 
plays an important role in the construction of the ultralocal Lax pair. 
Such an interplay between integrability and the para-complex structure 
is also crucial in our analysis. Furthermore, the para-complex target spaces 
we shall consider are analogues of  complex target spaces considered 
by D. Bykov in \cite{Bykov:2017vsm,Bykov:2016rdv,Bykov:2019jbz} (see also 
\cite{Bykov:2014efa,Bykov:2015pka}). 
The reason why we depart from the case of target spaces having a complex 
 structure is the following. For complex $\mathbb{Z}_T$-cosets with $T>2$ and  a 
 worldsheet with Minkowski signature,  one would encounter known (see for instance 
  \cite{Young:2005jv}) problems 
 with reality conditions already at the level of the action. 
 Furthermore,  even when $T=2$, the construction of the ultralocal Lax 
 connection for complex target spaces would spoil reality conditions.
Let us note that such problems have already been pointed out in \cite{Bazhanov:2017nzh} for the ultralocal Lax connection 
 of the O(3) non-linear $\sigma$-model considered there. This is the reason why each Lie 
 algebra we shall consider is the split real 
form of a complex Lie algebra  and why we shall deal with para-complex instead 
of complex cosets. 
 
\medskip

The plan of this article is the following.  In section \ref{sec kaehler}, we describe the para-complex 
$\mathbb{Z}_T$-cosets $G/H$ we shall consider. Their para-complex structure 
 and the 
$\mathbb{Z}_T$-grading are both defined from a particular element of the Lie 
algebra $\g$ of $G$. We explain how these three characteristics are related to each other. 

We proceed in section \ref{sec: Lag analysis} with the Lagrangian analysis. 
We first explain how the action of generic $\mathbb{Z}_T$-cosets  
may be greatly simplified, in the case of para-complex $\mathbb{Z}_T$-cosets, by 
adding to it a total derivative.  The main advantage of such a procedure is that 
it enables to find easily a conserved and gauge-invariant current $\K_\pm$, which is also flat.
 This current is associated with the isometry 
of the para-complex $\mathbb{Z}_T$-cosets.  The existence 
of this conserved and flat current allows one to define a Lax connection, $\L_\pm$, 
which is of the Zakharov-Mikhailov \cite{Zakharov:1973pp} type. 
One important property of this Lax connection is its gauge invariance, 
which is inherited from that of the current. 
 We end this section by explaining how the Lax connection  $\L_\pm$ is related to 
 the ordinary Lax connection of $\mathbb{Z}_T$-coset $\sigma$-models by a formal 
 gauge transformation depending on the spectral parameter. 
 Section \ref{sec: Lag analysis} generalises results obtained 
 in \cite{Bykov:2017vsm,Bykov:2016rdv,Bykov:2019jbz} for some complex target spaces. 
 
 Section \ref{sec: Ham analysis} is devoted to the Hamiltonian analysis. 
 We start by giving the canonical expression of the conserved and flat current. 
 Since the action admits a gauge symmetry, we recall that there is a freedom 
 to add to the Hamiltonian expression  of any quantity a 
 term proportional to the first-class constraint 
 associated with the gauge invariance. We explain how we use this freedom  
 in order to have a strongly vanishing Poisson bracket between $\K_+$ and $\K_-$.  
 We also give details of the computation of the Poisson bracket of $\K_\pm$ with itself. All these Poisson brackets are ultralocal. It is then immediate that 
 the Poisson 
 brackets of the Lax connection are ultralocal. Furthermore, they take the standard $R$-matrix form. 
 This implies that the monodromy 
 matrix satisfies a Poisson algebra which is the classical analogue of a Yangian.
  Finally, we make some comments in the conclusion.

\section{Para-complex \texorpdfstring{$\mathbb{Z}_T$-cosets}{ZT-cosets}}
\label{sec kaehler}

In this section, we describe the particular class of $\mathbb{Z}_T$-cosets 
which we shall consider. Let $G$ be a 
semisimple real Lie group whose Lie  algebra $\g$ is assumed to be   
the split real form of a complex Lie algebra $\g^\mathbb{C}$.

\paragraph{The $\mathbb{Z}$-gradation.}

An important role in the whole analysis is played by an element $\uucheche$ in 
the Cartan subalgebra of $\g$ whose eigenvalues in the adjoint representation are integers between $-T+1$ and $T-1$. 
This defines a $\mathbb{Z}$-gradation
\beq
\g =\bigoplus_{k=-T+1}^{T-1} \g^{[k]} ,\label{zgrading}
\eeq
where $\g^{[k]}$ is the eigenspace of $\ad_\uucheche$ corresponding
to the eigenvalue $k$ with $-T<k<T$. Note that this $\mathbb{Z}$-gradation is not 
cyclic. In particular, we have
\beq 
\forall m\in \g^{[k]},\,\,\forall n\in \g^{[k']},\quad[m,n] = 0 \quad \mbox{if} \quad |k + k'| \geqslant T. \label{toobig}
\eeq

\paragraph{The $\mathbb{Z}_T$-gradation.}

Before explaining how to construct the distinguished element $\uucheche$, let us first 
describe how it also induces a $\mathbb{Z}_T$-gradation on $\g$. Let  
$\omega = e^{ 2 i \pi /T}$ 
and define the automorphism $\sigma$ of $\g^\mathbb{C}$ by  
\beq 
\sigma = \omega^{\ad_\uucheche}
= \exp\Bigl( \frac{2 i \pi}{T} \ad_\uucheche \Bigr).
\label{sigma on fk spaces}
\eeq
This is, by construction, an automorphism of order $T$. It defines a $\mathbb{Z}_T$-gradation
\beq 
\g = \bigoplus_{k=0}^{T-1} \g^{(k)} \label{ZTgrading}
\eeq
of the Lie algebra $\g$, where $\g^{(k)}$ is the eigenspace of $\sigma$ corresponding to the eigenvalue $\omega^k$.
In particular, we have
\begin{equation*}
\forall m\in \g^{(k)},\,\,\forall n\in \g^{(k')},\quad[m,n] \in \g^{(k+k' \; \textup{mod}\, T)}\,
\end{equation*}
for any $k, k' = 0, \ldots, T-1$, which is to be compared with \eqref{toobig} for the 
$\mathbb{Z}$-gradation. In fact, by using the property that $\omega^T=1$, we see that the relation between the $\mathbb{Z}$-gradation \eqref{zgrading} and the $\mathbb{Z}_T$-gradation \eqref{ZTgrading} is
\beq \label{Zt vs Z}
\g^{(0)} = \g^{[0]} \qquad \mbox{and} \qquad \g^{(k)} = \g^{[k]} \oplus \g^{[-T+k]}.
\eeq
We shall decompose any $m^{(k)} \in \g^{(k)}$, using the direct sum decomposition \eqref{Zt vs Z}, as 
\beqz
m^{(k)} = m^{[k]} + m^{[-T+k]}
\eeqz
with $m^{[k]} \in \g^{[k]}$ and $m^{[-T+k]}\in \g^{[-T+k]}$.

Let us introduce the notation $\h \equiv \g^{(0)} = \g^{[0]}$. The subgroup $H$ of $G$ 
with Lie algebra $\h$ is the centralizer of $\uucheche$ under the adjoint action of $G$. Note that 
$H$ has a 
non-trivial center, which contains at least the abelian subgroup of $G$ generated by $\uucheche$.

\paragraph{Para-complex structure.}

For any element $Y=\sum_{k=-T+1}^{T-1}Y^{[k]}$ of the Lie algebra $\g$, it will be convenient to use the notations
\beqz
Y^<=P^<(Y)=\sum_{k=-T+1}^{-1}Y^{[k]},\qquad Y^>=P^>(Y)=\sum_{k=1}^{T-1}Y^{[k]},\qquad 
Y^\geqslant=P^\geqslant(Y)=\sum_{k=0}^{T-1}Y^{[k]},
\eeqz
where $P^<$, $P^>$ and $P^\geqslant$ are projectors on the subalgebras of $\g$ with 
respectively negative, positive and non-negative grades. We denote by 
$\g^<$ and $\g^>$ the images of $P^<$ and $P^>$. 

Let us then define the map $J=P^<-P^>$ acting on $\g$. Its restriction to   
$\g^< \oplus \g^>$ satisfies 
the two properties
\begin{gather*}
J^2(X) =\id, \\
[J(X),J(Y)]-J([X,J(Y)]+[J(X),Y])+[X,Y] =0,
\end{gather*}
for any $X, Y \in \g^< \oplus \g^>$. The latter equation may be interpreted as the vanishing of the Nijenhuis tensor 
associated with $J$,  which means that $J$ defines a para-complex structure on 
$G/H$ \cite{Libermannphdthesis}.

\paragraph{Construction of $\uucheche$.}

The distinguished element $\uucheche$ which defines the $\ZZ$-gradation 
in \eqref{zgrading} and the para-complex structure may be constructed as follows. Let $\{\alpha_i \}_{i=1}^l$ denote a set of positive simple roots of 
the Lie algebra $\g$. The longest positive root is $\theta=\sum_{i=1}^l a_i\alpha_i$, where $a_i$ are positive integers. 
We denote by $\{\check{\omega}_i\}_{i=1}^l$ the basis of the Cartan subalgebra  of $\g$ formed of 
fundamental co-weights defined by
$\alpha_j(\check{\omega}_i)=\delta_{ij}$. 
We then choose
\beqz
\uucheche=\sum_{i=1}^l b_i\check{\omega}_i,
\eeqz
where $b_i$ are non-negative integers to be fixed shortly. If $\alpha=\sum_{i=1}^l  m_i\alpha_i$ is a 
positive root, with $E_\alpha$, $F_\alpha$ denoting the corresponding root vectors in $\g$, then
\beqz
[\uucheche, E_\alpha]=\bigg(\sum_{i=1}^l  b_im_i\bigg)E_\alpha,
\qquad [\uucheche, F_\alpha]=-\bigg(\sum_{i=1}^l  b_im_i\bigg)F_\alpha.
\eeqz
We shall therefore fix the $b_i$ by requiring that $T-1=\sum_{i=1}^l  b_i a_i$.
Let $N_0 \subset \{1,\ldots, l\}$ be such that $b_i=0$ if and only if $i\in N_0$. We then have 
that $E_\alpha, F_\alpha \in \g^{[0]}=\h$ whenever the root $\alpha$ is of the form $\alpha = \sum_{i\in N_0} m_i\alpha_i$. Notice that for a generic choice of the $b_i$'s, some of the subspaces $\g^{[k]}$ may be trivial. 
 
Let us finally note    that the definition of the $\mathbb{Z}_T$-automorphism in 
\eqref{sigma on fk spaces} is such that the root vector associated with the negative 
of the longest root has grade $1$ with respect to the $\mathbb{Z}_T$-gradation, namely
\beqz
[\uucheche,F_\theta]=(1-T)F_\theta\,\,\Longrightarrow \sigma(F_\theta)=\omega F_\theta.
\eeqz

\paragraph{Decomposition of the quadratic Casimir.}

Let $\{ I_a \}$ be a basis of $\g$ and $\{ I^a\}$ be its dual basis with respect to 
the opposite of the Killing form $\kappa$.
The ad-invariance of $\kappa$ implies that $\kappa(m^{[k]}, m^{[p]})=0$ 
unless $k=-p$. This implies that the subalgebras $\g^>$ and $\g^<$ of $\g$ are isotropic. 

Let us also fix a basis $\{ I_a^{[k]} \}$ of $\g^{[k]}$, for 
each $k = -T+1, \ldots, T-1$, and let $\{ I^{a [-k]} \}$ denote its dual basis. 
A basis of $\g^{(0)} = \g^{[0]}$ is then given by $\{ I_a^{(0)} \} = \{ I_a^{[0]} \}$ 
and its dual basis is given by $\{ I^{a (0)} \} = \{ I^{a [0]} \}$. 
The quadratic Casimir can be written as
\beqz
C\ti{12} = \sum_a I_a \otimes I^a = C\ti{12}^{<>} + C\ti{12}^{\geqslant \leqslant},
\eeqz
with
\beqz 
C\ti{12}^{<>} =  \sum_{k=-T+1}^{-1} \sum_a  I_a^{[k]} \otimes I^{a [-k]} 
\qquad \mbox{and} \qquad C\ti{12}^{\geqslant \leqslant} 
=   \sum_{k=0}^{T-1} \sum_a I_a^{[k]} \otimes I^{a [-k]}.
\eeqz
  
\paragraph{Examples.}

In the case $T=2$, one could have relaxed the condition that $\g$ is the split real form of 
a complex Lie algebra. For compact real forms, the $\mathbb{Z}_2$-cosets 
constructed in the previous paragraphs correspond  
to  K\"ahlerian symmetric spaces. These are the cosets considered 
in \cite{Brodbeck:1999ib}. We shall, however, 
not consider these cases because their ultralocal Lax connection is not compatible
with reality conditions. The reason 
for this may be illustrated  
in the case of the coset SU(3)/(SU(2)  
$\!\times\!$ U(1)) $\simeq \CC \PP^2$. Indeed, taking $\uucheche= \mbox{diag}(\frac{1}{3},\frac{1}{3},-\frac{2}{3})$ 
is fine in order for the subalgebra $\mathfrak{su}(2) \oplus \mathfrak{u}(1)$ 
to correspond to the eigenspace of the adjoint action of 
$\uucheche$ with null eigenvalue. However, it is then clear that the two other eigenspaces 
are not subspaces of $\mathfrak{su}(3)$.  
 
\medskip

For $T \geqslant 2$, and to fix the ideas,  the pseudo-Riemannian manifolds 
such as 
\beqz
\frac{\text{SL}(p_1 + \cdots + p_T)}
{\text{S}( 
\text{GL}(p_1)\times \cdots \times \text{GL}(p_T))}
\eeqz
are para-complex $\mathbb{Z}_T$-cosets and non-symmetric whenever $T>2$.

\section{Lagrangian analysis}
\label{sec: Lag analysis}

\subsection{Action}

We start with the action \cite{Young:2005jv} of $\mathbb{Z}_T$-cosets,
\beq \label{action ZT}
S[g] = K \iint \dd x^+ \, \dd x^- \; \sum_{k=1}^{T-1} k \,  \kappa(j_+^{(k)}, j_-^{(T-k)}).
\eeq
The field $g(x,t)$ takes values in the Lie group $G$ and  $j_\pm = g^{-1} \partial_\pm g$ 
with $x^\pm = \ha(t\pm x)$ 
and $\partial_\pm = \partial_t \pm \partial_x$. The target space is the coset $G/H$
since the action is invariant under the gauge transformation 
\beq \label{gauge transfo}
g(x,t) \longmapsto g(x,t) h(x,t)
\eeq
with $h(x,t)$ taking values in $H$. 

A short computation shows that the action (\ref{action ZT}) may be rewritten in 
terms of the $\mathbb{Z}$-graded components of the current $j_\pm$ as
\beq \label{action ZT in Z gradation}
S[g] = K \iint \dd x^+ \, \dd x^- \; \sum_{k=1}^{T-1} \Bigl( k \,  \kappa(j_+^{[k]}, j_-^{[-k]})
+(T-k) \, \kappa(j_+^{[-k]}, j_-^{[k]}) \Bigr).
\eeq
It may further be separated into a metric part and a $B$-field part as follows
\begin{align} \label{action ZT G and B}
S[g] = K \iint \dd x^+ \, \dd x^- \; \sum_{k=1}^{T-1} & \biggl(\frac{T}{2} \Bigl(\kappa(j_+^{[k]}, j_-^{[-k]})+
\kappa(j_+^{[-k]}, j_-^{[k]})\Bigr)\\
&+\frac{2k-T}{2}\Bigl(\kappa(j_+^{[k]}, j_-^{[-k]})- \kappa(j_+^{[-k]}, j_-^{[k]})\Bigr)\biggr). \nonumber
\end{align}
Aside from the fact that the grade zero is absent, the metric part is clearly independent of the 
$\mathbb{Z}$-gradation. Indeed, two $\mathbb{Z}$-gradations with the same 
zero grade component $\g^{[0]}$ give the same 
metric. The $B$-field part may, at first sight, seem
to depend on it. However, using the Maurer-Cartan equations, invariance of the Killing form 
and  the definition of the $\mathbb{Z}$-gradation one has 
\beqz
\kappa(\uucheche,\partial_-j_+-\partial_+ j_-)=\sum_{k=1}^{T-1}k\left(\kappa(j_+^{[k]}, j_-^{[-k]})- \kappa(j_+^{[-k]}, j_-^{[k]})\right).
\eeqz
Thus, the term in the $B$-field proportional to $k$ is in fact a total derivative. 
This means that the $\sigma$-model may be defined by the action
\beq \label{action simplified}
S[g] = K T \iint \dd x^+ \, \dd x^- \; \sum_{k=1}^{T-1} \kappa(j_+^{[-k]}, j_-^{[k]})=
KT\int\dd x^+\,\dd x^-\;\kappa(j_+^<,j_-^>).
\eeq
The $B$-field part of the action (\ref{action simplified}) may simply be written as
\beqz  
\frac{K T}{2} \iint \dd x^+ \, \dd x^- \; \kappa(j_+,J(j_-)).
\eeqz
This is fully analogous, in the split framework, to the models considered in the compact 
case in \cite{Brodbeck:1999ib} and in \cite{Bykov:2017vsm,Bykov:2016rdv}.
 
\subsection{Flat and conserved current}

The action \eqref{action simplified} is invariant under the global symmetry $g(x,t) \to 
g_0 \, g(x,t)$ with $g_0 \in G$. 
A conserved current $\K_\pm$ associated with this symmetry is obtained by 
applying Noether's theorem. Furthermore, the equations of motion 
correspond to the equation of conservation 
\beqz
\partial_+ \K_- + \partial_- \K_+ =0 
\eeqz
of this current whose light-cone components are given explicitly by
 \begin{align} \label{improved flat}
\K_+ =  - 2  
g j_+^{<} g^{-1}, \qquad
\K_- = - 2  
g j_-^> g^{-1}.
 \end{align}
The current $\K_\pm$ is also gauge-invariant. This is immediate since under 
 a gauge transformation \eqref{gauge transfo}, we 
have, for $k\neq 0$,  
\beqz
j_\pm^{[k]}(x,t)  \longmapsto h^{-1}(x,t) j_\pm^{[k]}(x,t) h(x,t).
\eeqz
The overall factor in this conserved current $\K_\pm$ has been fixed in order for it to also be flat, namely we have
\beqz 
\partial_+ \K_- - \partial_- \K_+ + [\K_+,\K_-] =0.
\eeqz
However, we postpone the proof of this flatness property until the next subsection, where 
we will establish this result in an indirect way. 

\medskip

The Noether current is not unique. In fact, starting from the action \eqref{action ZT}, 
one would have naturally found
\beqz 
K_+ = \sum_{k=1}^{T-1} k g j_+^{(k)} g^{-1} 
\qquad \mbox{and}
\qquad
K_- = \sum_{k=1}^{T-1} k g j_-^{(T-k)} g^{-1}. 
\eeqz
It is then clear from the analysis of the previous section that the existence of the element $\uucheche \in \g$ 
 allows one to introduce an improvement term 
  relating the two  currents 
\beqz
\K_\pm = -\frac{2}{T} \bigl( K_\pm \pm  \partial_\pm(g \uucheche g^{-1})\bigr).
\eeqz

\medskip

Let us note that for symmetric space $\sigma$-models, that is when $T=2$, 
the conserved current $K_\pm$ can be made flat after an overall re-scaling 
to $- 2 K_\pm$. However, for $T>2$, it is not possible to make the conserved current 
$K_\pm$ also be flat in this way. The existence of the  real, flat and conserved
current \eqref{improved flat} is thus a characteristic of para-complex 
$\mathbb{Z}_T$-cosets. Furthermore, as we shall prove in section \ref{sec: Ham analysis}, its 
Poisson brackets with itself are ultralocal. 

\subsection{Lax connection}

If one has a flat and conserved current,  one can define the  
Lax connection which is of 
Zakharov-Mikhailov \cite{Zakharov:1973pp} type, 
\beq \label{ZM UL LP}
\L_\pm(\lambda) = \frac{\K_\pm}{1\mp \lambda},
\eeq
where $\lambda$ denotes the spectral parameter. 
This Lax connection is flat on-shell, \emph{i.e.} the conservation and flatness of $\K_\pm$ is equivalent to the zero-curvature equation
\beq  \label{zce Lul}
\partial_+ \L_-(\lambda) - \partial_- \L_+(\lambda) + [\L_+(\lambda),\L_-(\lambda)] = 0.
\eeq
Let us discuss a few simple properties of this Lax connection before 
showing, in section \ref{sec: Ham analysis}, the ultralocality of 
its Poisson brackets. 

\paragraph{Gauge invariance.} 
A crucial property of $\L_\pm(\lambda)$ is its gauge invariance. 
This follows from the gauge invariance of the current itself. 
 This 
property has a very important consequence at the Hamiltonian level. Indeed, 
when the gauge invariance is fixed, Poisson brackets have to be replaced by Dirac brackets. 
However, the Dirac bracket of two gauge invariant quantities is equal to their Poisson 
bracket
(see for instance \cite{henneauxteitelboim_1994}). This 
implies that the ultralocal structure computed in the next section is unchanged  
when the gauge 
invariance  is fixed. 

\paragraph{Link with the ordinary Lax connection of $\mathbb{Z}_T$-cosets.}

The ordinary Lax connection $\LO_\pm(z)$ of $\mathbb{Z}_T$-cosets is \cite{Young:2005jv}
\begin{equation} \label{Lpm lagrangian}
\LO_+(z) = \sum_{k=0}^{T-1} z^{k} j_+^{(k)}, \qquad
\LO_-(z) = \sum_{k=0}^{T-1} z^{-k} j_-^{(T-k)},
\end{equation}
where the spectral parameter is denoted here by $z$. 
Let us then define
\beqz 
\alpha(z) = \exp(\uucheche \ln z ),
\eeqz
which is valued in $G^\CC$.  
It satisfies the property 
\beqz 
\alpha(z)^{-1} m \alpha(z)= z^{-k} m, \qquad \forall m \in \g^{[k]} 
\eeqz
for every $k = -T+1, \ldots, T-1$. Recall that the zero curvature equation \eqref{zce Lul} is invariant under formal gauge 
transformations. We apply the formal gauge transformation 
\beq \label{Eq: gauge transfo L}
\LO_\pm^U(z) = U(z) \LO_\pm(z) U(z)^{-1} + U(z) \partial_\pm U(z)^{-1} 
\eeq
depending on the spectral parameter $z$, where 
\beqz
U(z,x,t) =  g(x,t) \, \alpha(z)^{-1}.
\eeqz
Let us work out the expression for the gauge transformed Lax connection 
$\LO^U_\pm(z)$. We first observe that  
\beq \label{inho}
U \partial_\pm U ^{-1} =   - g j_\pm g^{-1}
\eeq
and $U \LO_\pm U^{-1} = g \, \bigl(\alpha^{-1} \LO_\pm   \alpha\bigr) \, g^{-1}$. 
Focusing on $\LO_+(z)$, we obtain successively: 
\begin{align}
\alpha(z)^{-1} \LO_+(z) \alpha(z) &= \sum_{k=0}^{T-1} z^{k} \alpha(z)^{-1}  j_+^{(k)}
\alpha(z) \nonumber \\
&= \alpha(z)^{-1}  j_+^{(0)} \alpha(z) + 
\sum_{k=1}^{T-1} z^{k} \alpha(z)^{-1} \bigl(
j_+^{[k]}+ j_+^{[k-T]}
\bigr)
\alpha(z) \nonumber \\
&=  j_+^{(0)} +\sum_{k=1}^{T-1} \Bigl( 
j_+^{[k]} + z^{T} j_+^{[k-T]} \label{uLpuinv}
\Bigr).
\end{align}
It therefore follows from \eqref{Eq: gauge transfo L}, \eqref{inho} and \eqref{uLpuinv} 
that
\beqz
 \LO_+^U(z) = g \biggl(j_+^{(0)} +\sum_{k=1}^{T-1} \Bigl( 
j_+^{[k]} + z^{T} j_+^{[k-T]} \Bigr) - j_+ \biggr) g^{-1}=
(z^T-1) \sum_{k=1}^{T-1}   g \,  j_+^{[k-T]}   g^{-1}.
\eeqz
Proceeding in the same way for $\LO_-(z)$, 
and recalling the expressions \eqref{improved flat} of $\K_\pm$, 
we obtain
\beq \label{ulax lag expr}
 \LO_\pm^U(z) = - \ha (z^{\pm T}-1) \K_\pm.
\eeq
Finally, performing also the following change of spectral parameter
\beqz
\lambda \longmapsto z(\lambda) = \left( 
\frac{\lambda + 1}{ \lambda - 1}
\right)^{1/T},
\eeqz
we arrive at the relation
\beqz 
\LO^U_\pm\big( z(\lambda) \big) = \L_\pm(\lambda). 
\eeqz
In other words, the ultralocal Lax connection $\L_\pm(\lambda)$  coincides, up to a change of spectral parameter, with  
a formal gauge transformation of the ordinary Lax connection $\LO_\pm(z)$. 
An immediate consequence of this is that the Lax connection $\L_\pm(\lambda)$ is flat, since we know that $\LO_\pm(z)$ is flat and that formal gauge transformations preserve the flatness 
property. Moreover, since $\L_\pm(\lambda)$ is of the Zakharov-Mikhailov form, this proves indirectly that the current $\K_\pm$ is also flat on-shell. 
 
\section{Hamiltonian analysis and ultralocality}
\label{sec: Ham analysis}

\subsection{Result of the canonical analysis}

The phase space is parameterised by fields $g(x)$ and $X(x)$ taking
values in $G$ and $\g$, respectively, the pair of which describes a field 
valued in the cotangent bundle $T^\ast G$.  They satisfy the canonical Poisson brackets,
which written in tensorial notation read
\begin{subequations}\label{Eq:PBTstarG}
\begin{align}
\left\lbrace g\ti{1}(x), g\ti{2}(y) \right\rbrace & = 0, \\
\left\lbrace X\ti{1}(x), g\ti{2}(y) \right\rbrace & = g\ti{2}(x) C\ti{12} \delta_{xy},\label{Eq:PBXg} \\
\left\lbrace X\ti{1}(x), X\ti{2}(y) \right\rbrace & = -
\left[ C\ti{12}, X\ti{2}(x) \right] \delta_{xy}.\label{Eq:PBXX}
\end{align}
\end{subequations}
The canonical analysis associated with the action \eqref{action simplified} is standard. 
We shall not reproduce its details here, which lead to the following 
relation 
\beqz 
X=\frac{KT}{2}(j_-^> + j_+^< ).
\eeqz
There is  a first-class constraint $X^{[0]}=0$, which corresponds to the gauge invariance \eqref{gauge transfo} of the action.

\medskip

Using (\ref{improved flat}), we immediately obtain the phase space expressions of the flat current:
\beq \label{Kpm new exp X}
\K_+ = - \frac{4}{KT} gX^<g^{-1},\qquad \K_- = - \frac{4}{KT} gX^\geqslant g^{-1}.
\eeq
Note that here we have added to $\K_-$ the extra term $- \frac{4}{K T} g X^{[0]} g^{-1}$ which is proportional 
to the constraint. This freedom to add terms proportional to the constraint  
is a standard procedure in integrable field theories with gauge symmetry  
(see \cite{Magro:2008dv,Vicedo:2009sn}).
Indeed, as we shall see, the coefficient of this extra term has been fixed in order for the Poisson bracket between 
$\K_+$ and $\K_-$ to vanish strongly, that is without making use of the constraint. Note, however, that the chosen value of this coefficient also makes sense for the following reason. Let us consider the temporal component of the current $\K_\pm$, namely
\beqz 
\ha(\K_+ +\K_-) = - \frac{2}{K T} g X g^{-1}.
\eeqz
For any $\epsilon \in \g$ we then have 
\beqz 
\left\{ 
- \frac{K T}{2} \int \dd y \; \kappa\Bigl(\epsilon , \ha(\K_+(y) +\K_-(y))\Bigr), g(x) 
\right\}
= \epsilon g(x). 
\eeqz
This is what we expect in order for the time component of the current to 
generate the symmetries corresponding to left multiplication on $g$.

\subsection{Computation of the Poisson brackets of \texorpdfstring{$\K_\pm$}{Kpm}}

\paragraph{Ultralocality.}

A key property of the expression \eqref{Kpm new exp X} of the current is that 
it depends on the fields $X$ and $g$, but not on their spatial derivatives. This ensures {\em de facto}
that all Poisson brackets of the current, and thus of the Lax pair (\ref{ZM UL LP}), are ultralocal! 
This property alone explains why the para-complex $\mathbb{Z}_T$-cosets are so special. Indeed, the fact that there is 
no spatial derivative in \eqref{Kpm new exp X}  is  a consequence of the 
form of the simplified action \eqref{action simplified}.
 
\medskip

In the remainder of this subsection we compute all the Poisson brackets of the current $\K_\pm$. 
It is clear that they take the following form,
\beq \label{PB Ka Kb gen}
\{{\K_a}\ti{1}(x),{\K_b}\ti{2}(x')\} =  \frac{16}{K^2 T^2}
g\ti{1}(x)g\ti{2}(x') \, \alpha_{ab}(x) \,  g\ti{1}^{-1}(x)g\ti{2}^{-1}(x') \delta_{xx'},
\eeq
where $\alpha_{ab}$ belongs to the tensor product of two copies of the 
Lie algebra $\g$, and $a, b = `\pm\text{'}$. Since the Poisson brackets are ultralocal, we shall not indicate the spatial dependence in intermediate 
computations.  
Each  $\alpha_{ab}$ is the sum of three terms:
\begin{align}
\alpha_{ab} &= g\ti{1}^{-1} g\ti{2}^{-1} \{ g\ti{1} P\ti{1}^{s(a)} X\ti{1} g\ti{1}^{-1},  
g\ti{2} P\ti{2}^{s(b)} X\ti{2} g\ti{2}^{-1} \}   g\ti{1} g\ti{2} \nonumber \\
&=P\ti{1}^{s(a)} P\ti{2}^{s(b)} \{ X\ti{1}, X\ti{2} \} 
+ P\ti{2}^{s(b)} \bigl[ 
g\ti{1}^{-1} \{ g\ti{1}, X\ti{2} \}, P\ti{1}^{s(a)} X\ti{1}
\bigr]
+ P\ti{1}^{s(a)} \bigl[ 
g\ti{2}^{-1} \{ X\ti{1} , g\ti{2}\}, P\ti{2}^{s(b)} X\ti{2}
\bigr]
\nonumber \\
&=  - P\ti{1}^{s(a)} P\ti{2}^{s(b)} [C\ti{12},X\ti{2}] + P\ti{2}^{s(b)} [C\ti{12}, P\ti{2}^{s(a)} X\ti{2}] 
 + [P\ti{1}^{s(a)} C\ti{12}, P\ti{2}^{s(b)} X\ti{2}], \label{result for alab}
\end{align}
where $s(+) =\, `<\text{'}$ and $s(-)=\, `\geqslant\text{'}$. To establish this result, we have made use of \eqref{Eq:PBXX}, \eqref{Eq:PBXg}, the 
antisymmetry of the Poisson bracket and the identity $[C\ti{12}, M\ti{1} + M\ti{2}]=0$ valid for 
any $M \in \g$. It remains then to compute $\alpha_{ab}$ for each possibility. 

\paragraph{Poisson bracket $\{\K_+,\K_-\}$.}
In this case, $a=`+\text{'}$, $b=`-\text{'}$ and thus $s(a)= \, `<\text{'}$ and $s(b)= \, `\geqslant\text{'}$. We have therefore
\beq \label{alpha +-}
\alpha_{+-}=  -P\ti{2}^\geqslant [C\ti{12}^{<>}, X\ti{2}] + P\ti{2}^\geqslant [C\ti{12}, X\ti{2}^<] 
+ [C\ti{12}^{<>}, X\ti{2}^\geqslant].
\eeq
For the second term in the r.h.s. of \eqref{alpha +-}, the projector $P\ti{2}^\geqslant$ 
forces the grading of the 
commutator in the second tensorial space to be greater than or equal to zero. However,  since the 
grading of $X^<\ti{2}$ is negative, we have:
\beqz 
P\ti{2}^\geqslant [C\ti{12}, X\ti{2}^<]  =  P\ti{2}^\geqslant [C\ti{12}^{< >}, X\ti{2}^<].
\eeqz
The grading  in the second tensorial space of the third term in the r.h.s. of \eqref{alpha +-} is strictly 
positive. One has therefore the identity 
\beqz 
[C\ti{12}^{<>}, X\ti{2}^\geqslant] = P\ti{2}^\geqslant[C\ti{12}^{<>}, X\ti{2}^\geqslant].
\eeqz
It is then clear that the sum \eqref{alpha +-} vanishes, and thus that 
  $\{{\K_+}\ti{1}(x),{\K_-}\ti{2}(x')\} =0$.

\paragraph{Poisson bracket $\{\K_-,\K_-\}$.}

In this case, $a=b=`-\text{'}$ and thus $s(a)=s(b)=\, `\geqslant\text{'}$. We proceed 
in the same way as for the previous computation. We obtain:
\begin{align*}
\alpha_{--}&=  -P\ti{2}^\geqslant [C\ti{12}^{\geqslant \leqslant}, X\ti{2}] + P\ti{2}^\geqslant [C\ti{12}, X\ti{2}^\geqslant] 
+ [C\ti{12}^{\geqslant \leqslant}, X\ti{2}^\geqslant]
\\
&=-P\ti{2}^\geqslant [C\ti{12}^{\geqslant \leqslant}, X\ti{2}^\geqslant] + P\ti{2}^\geqslant [C\ti{12}, X\ti{2}^\geqslant] 
+ (P\ti{2}^< + P\ti{2}^\geqslant) [C\ti{12}^{\geqslant \leqslant}, X\ti{2}^\geqslant]\\
&=  P\ti{2}^\geqslant [C\ti{12}, X\ti{2}^\geqslant] 
+ P\ti{2}^< [C\ti{12}^{\geqslant \leqslant}, X\ti{2}^\geqslant]=  P\ti{2}^\geqslant [C\ti{12}, X\ti{2}^\geqslant] 
+ P\ti{2}^< [C\ti{12}, X\ti{2}^\geqslant]
=  [C\ti{12}, X\ti{2}^\geqslant].
\end{align*}
To conclude the computation, we use
the property $g\ti{1}(x)g\ti{2}(x)  C\ti{12} g\ti{1}^{-1}(x)g\ti{2}^{-1}(x)  =C\ti{12}$ and obtain 
$\{{\K_-}\ti{1}(x),{\K_-}\ti{2}(x')\} =- \frac{4}{K T}
 [C\ti{12} ,  {\K_-}\ti{2}(x)]
  \delta_{xx'}$.

\paragraph{Poisson bracket $\{\K_+,\K_+\}$.}

In this case, $a=b=`+\text{'}$ and thus $s(a)=s(b)=`<\text{'}$.  
There are only minor differences with the previous computation since we obtain 
\begin{align*}
\alpha_{++}&=  -P\ti{2}^< [C\ti{12}^{< >}, X\ti{2}] + P\ti{2}^< [C\ti{12}, X\ti{2}^<] 
+ [C\ti{12}^{<>}, X\ti{2}^<]
\\
&=  -P\ti{2}^< [C\ti{12}^{< >}, X\ti{2}^<] + P\ti{2}^< [C\ti{12}, X\ti{2}^<] 
+ [C\ti{12}^{<>}, X\ti{2}^<]
=   P\ti{2}^< [C\ti{12}, X\ti{2}^<] 
+  P\ti{2}^\geqslant [C\ti{12}^{<>}, X\ti{2}^<]\\
&=   P\ti{2}^< [C\ti{12}, X\ti{2}^<] 
+  P\ti{2}^\geqslant [C\ti{12}, X\ti{2}^<]
=   [C\ti{12}, X\ti{2}^<].
\end{align*}
This gives the last Poisson bracket, $\{{\K_+}\ti{1}(x),{\K_+}\ti{2}(x')\} =- \frac{4}{K T}
 [C\ti{12} ,  {\K_+}\ti{2}(x)]
  \delta_{xx'}$. 
  
In conclusion, we have shown that
\begin{subequations} \label{PB K K all}
\begin{align}
\{{\K_+}\ti{1}(x),
{\K_-}\ti{2}(x')\} &=0, \label{PB Kp Km}\\
\{{\K_\pm}\ti{1}(x),
{\K_\pm}\ti{2}(x')\} &= - \frac{4}{K T} [C\ti{12}, {\K_\pm}\ti{2}(x)] \delta_{xx'}. \label{PB Kpm Kpm}
\end{align}
\end{subequations}

\subsection{Poisson brackets of the Lax connection and Yangian}
 
It is then straightforward to compute all the Poisson brackets of the Lax connection from 
its definition in \eqref{ZM UL LP} and the above Poisson brackets \eqref{PB K K all}. The result is:
\begin{subequations} \label{PB L L all}
\begin{align}
\{{\L_+}\ti{1}(\lambda,x),{\L_-}\ti{2}(\mu,x')\} &=0, \label{PB Lp Lm}\\
\{{\L_\pm}\ti{1}(\lambda,x),{\L_\pm}\ti{2}(\mu,x')\} &= \mp \frac{4}{K T} 
\left[\frac{C\ti{12}}{\mu - \lambda}, {\L_\pm}\ti{1}(\lambda,x) + 
{\L_\pm}\ti{2}(\mu,x)\right] \delta_{xx'}. \label{PB Lpm Lpm}
\end{align}
\end{subequations}

The Poisson bracket of the Lax matrix $\L = \ha(\L_+ -\L_-)$ is then
\beq
\{{\L}\ti{1}(\lambda,x),{\L}\ti{2}(\mu,x')\} = - \frac{2}{K T} 
\left[\frac{C\ti{12}}{\mu - \lambda}, {\L}\ti{1}(\lambda,x) + 
{\L}\ti{2}(\mu,x)\right] \delta_{xx'}. \label{PB L L}
\eeq
We then define the monodromy 
\beqz
T(\lambda) = P \overleftarrow{\exp} \Bigl( - \int_W  \dd x \,\, \L(\lambda,x) \Bigr)
\eeqz
where $W$ is either the circle $S^1$ or $\mathbb{R}$. It is a consequence of the zero-curvature equation \eqref{zce Lul} that $T(\lambda)$ is conserved in time when $W = \mathbb{R}$, provided the Lax connection decays sufficiently fast at $\pm \infty$, or that the invariants of $T(\lambda)$ are conserved in time when $W = S^1$. In particular, this provides an indirect proof that the monodromy matrix $T(\lambda)$ (or rather its invariants in the case $W = S^1$) Poisson commutes with the Hamiltonian.

The Poisson brackets of the monodromy take \cite{Izergin:1980pe} 
 the form of a Poisson algebra corresponding to a Yangian 
(see also \cite{deVega:1983gn,MacKay:1992he} and the reviews 
\cite{Bernard:1992ya,MacKay:2004tc,Loebbert:2016cdm}),
\beqz 
\{ T\ti{1}(\lambda) , T\ti{2}(\mu) \} = \frac{2}{KT} \bigg[ \frac{C\ti{12}}{\mu-\lambda}, T\ti{1}(\lambda) 
T\ti{2}(\mu) \bigg].
\eeqz
Note that because of the ultralocality of the Poisson bracket \eqref{PB L L}, there is 
no ambiguity in the computation of this Poisson algebra. 

\section{Conclusion}

We have shown that classical integrable $\sigma$-models 
 on para-complex $\mathbb{Z}_T$-coset
target spaces admit an ultralocal Lax connection, which is related to the standard 
one by a spectral parameter dependent formal gauge transformation. The most important 
open problem relating to this class of models is therefore to apply the Quantum Inverse 
Scattering Method to them. A natural related question is also to determine whether the approach 
developed by V. Bazhanov, G. Kotousov and S. Lukyanov  in \cite{Bazhanov:2017nzh} 
can be extended to integrable $\sigma$-models on para-complex $\mathbb{Z}_T$-coset target spaces.

One may also wonder if there are other integrable $\sigma$-models which admit an 
ultralocal Lax connection. 
For instance, there is \cite{Bazhanov:2017nzh}  a generalisation of the ultralocal Lax connection of  
the $O(3)$ non-linear $\sigma$-model for the  
sausage model, which is a deformation \cite{Fateev:1992tk} of the former.
It would therefore be interesting to investigate if the 
result of the present article could be extended to one-parameter 
deformations \cite{Delduc:2013fga,Hollowood:2014rla,Vicedo:2017cge} of para-complex $\mathbb{Z}_T$-cosets.

In general, the question of whether or not the classical integrability of a $\sigma$-model is 
preserved at the quantum level is a difficult one  
\cite{DAdda:1978vbw, Goldschmidt:1980wq,Abdalla:1980jt,Abdalla:1982yd,Evans:2004mu,Bykov:2019jbz,
Litvinov:2019rlv,
Fateev:2019xuq,Komatsu:2019hgc}. Having an ultralocal 
description of the para-complex analogues of certain problematic models like the $\CC \PP^N$ $\sigma$-model, with $N>1$, at the classical level is then quite appealing as 
it could provide another way to investigate the fate of their integrability at the quantum level.

Very recently, a general formalism for describing classical integrable field theories was proposed in \cite{Costello:2019tri}, which is based on a certain four-dimensional variant of Chern-Simons theory. In this setting, integrable field theories are constructed from the four-dimensional gauge theory by inserting different surface defects. 
In particular, it was shown in \cite{Costello:2019tri} that non-linear $\sigma$-models whose target space is a K\"ahler manifold can be constructed by using so-called order defects. The analysis of \cite{Vicedo:2019dej} suggests that such order defects can be used more generally to describe ultralocal integrable field theories. It would therefore be interesting to construct the class of $\sigma$-models with para-complex $\mathbb{Z}_T$-coset target spaces considered in the present paper within the framework of \cite{Costello:2019tri}.


\paragraph{Acknowledgments.} We thank B. Hoare and J.M. Maillet for useful discussions. This work is partially 
supported by the French Agence Nationale de la Recherche (ANR) under grant
 ANR-15-CE31-0006 DefIS. The work of S.L. is funded by the Deutsche Forschungsgemeinschaft (DFG, German Research Foundation) under 
 Germany's Excellence Strategy - EXC 2121 ``Quantum Universe'' - 390833306.

\providecommand{\href}[2]{#2}\begingroup\raggedright\endgroup

\end{document}